\documentclass[%
 reprint,
 amsmath,amssymb,
 aps,
sort&compress,% ← 添加这行，启用排序和压缩 +
numbers,        % ← 添加这行，强制使用数字引用 +
]{revtex4-2}

\usepackage{graphicx}% Include figure files
\usepackage{dcolumn}% Align table columns on decimal point
\usepackage{bm}% bold math
\usepackage{mathtools}  % 提供 \mathrlap
\usepackage{braket}  % 必须添加此宏包以支持 \ket 和 \bra 命令
\usepackage{float}
\usepackage{xcolor} 
\usepackage{subfigure}
\usepackage{booktabs} % 用于绘制高质量的三线表
\usepackage{multirow} % 用于合并单元格
\usepackage{siunitx}  % 用于数值列的对齐，可选
\usepackage{threeparttable} % <--- 新增此行以定义 tablenotes 环境
\usepackage{hyperref}
\hypersetup{
	colorlinks=true,        % 将链接文字显示为彩色，而不是带颜色的边框
	linkcolor=blue,         % 内部链接（章节、图表等）的颜色
	citecolor=blue,         % 引用标记的颜色
	urlcolor=blue,          % URL链接的颜色（适用于DOI）	
    pdfborder={0 0 0},  % 去除链接边框
    pdfstartview=FitH,   % 设置PDF打开视图
    pdfpagemode=UseOutlines,  % 显示书签
	allcolors=blue,      % <-- 将所有链接统一为蓝色
}
\setcitestyle{numbers}  % 强制使用数字引用格式（而非作者-年份）+
\allowdisplaybreaks  % +
\allowdisplaybreaks
\begin{document}

\preprint{APS/123-QED}

\title{Device-Independent Quantum Secret Sharing Protocol Enhanced by Advantage Distillation}% Force line breaks with \\

\author{Yong-Hui Yang\textsuperscript{1,2}}
\author{Jian-Hong Shi\textsuperscript{1,2}}\email{Contact author:shijianhong2011@163.com}
\author{Hong-Wei Li\textsuperscript{1,2}, Hai-Long Zhang\textsuperscript{1,2}, Yun-Teng Yang\textsuperscript{1,2}, Yu-Bing Zhu\textsuperscript{1,2}}
\author{Yan-Yang Zhou\textsuperscript{1,2}}\email{Contact author:zyy@qiclab.cn}

\affiliation{\textsuperscript{1}Henan Key Laboratory of Quantum Information and Cryptography, Zhengzhou, Henan 450000, China \\ \textsuperscript{2}Synergetic Innovation Center of Quantum Information and Quantum Physics, University of Science and Technology of China, Hefei 230026, China}

\date{\today}% It is always \today, today,

\begin{abstract}
Device-independent quantum secret sharing (DI-QSS) provides high security by eliminating the need to trust devices, yet its real performance is limited by channel loss and noise. This work extends advantage distillation from two-party quantum key distribution (QKD) to three-party DI-QSS, redesigning the corresponding data interaction and verification procedures. The technique is systematically applied to the basic protocol and three active improvement strategies: noise pre-processing, post-selection, and their combination. This approach enhances noise tolerance, reduces the required global detection efficiency threshold, and significantly extends the maximum secure communication distance. Numerical simulations demonstrate that for the basic protocol over fiber, the maximum secure distance increases from 0.16 km to 1.85 km, and the noise tolerance also improved from 10.17$\%$ to 28.49$\%$. The results show that generalizing advantage distillation to the three-party setting effectively strengthens the protocol’s robustness and practicality, enhancing its adaptability to practical noise and advancing the development of more reliable quantum secret sharing systems.
\end{abstract}

\maketitle

%\tableofcontents

\section{\label{sec:level1}Introduction}
Quantum secure communication\cite{zhou2023device,harney2022secure,senthoor2019communication,ying2024passive,zhang2017quantum}, founded on the fundamental principles of quantum mechanics, provides theoretically provable security for information transmission and has become a significant frontier in information security technology. However, its practical deployment and application still face numerous challenges, including practical issues such as device imperfections, channel loss, coherent attack\cite{xiao2025experimental,sheridan2010finite}, and eavesdropping attacks\cite{ekert1994eavesdropping,lee2022eavesdropping}. Currently, quantum key distribution (QKD)\cite{ekert1991quantum,lo2017measurement,xu2020secure,wang2022twin,zhang2022device,zapatero2023advances} technology has achieved substantial progress and is gradually transitioning to practical use, yet it is predominantly confined to two-party communication scenarios. As quantum networks expand in scale and complexity, the demand for multi-party collaborative secure communication is growing. Consequently, quantum secret sharing (QSS)\cite{hillery1999quantum,guo2003quantum,liao2014dynamic,fu2015long,zhang2024device,zhang2025device} technology, which enables the secure sharing and recovery of secrets among multiple participants, is seeing its research value and practical urgency become increasingly prominent.

The device-independent quantum secret sharing (DI-QSS)\cite{zhang2024device,zhang2025device} protocol represents an important advanced paradigm within quantum secret sharing. By certifying security through violations of quantum nonlocality\cite{popescu1994quantum,brunner2014bell}, it roots the protocol's security in the fundamental laws of quantum physics, eliminating the need for trusted modeling of the experimental apparatus. This approach effectively defends against attacks targeting actual device flaws and side-channels,  offering a very high level of theoretical security. However, the practical performance of existing DI-QSS protocols, especially those implemented using three-photon GHZ states, is severely constrained by photon loss and channel noise. This manifests as demanding requirements on the system's global detection efficiency and a low tolerance to quantum bit error rate (QBER), which limit their effective transmission distance and secret sharing rate in practical environments. To address the efficiency threshold requirements, various strategies have been explored in recent studies. These include optimized post-processing techniques\cite{ma2011improved}, the use of two-way classical communication\cite{tan2020advantage}, noise pre-processing\cite{ho2020noisy}, generalized Bell inequalities for enhanced security verification\cite{woodhead2021device,sekatski2021device,gonzales2021device}, complete statistical analysis based on von Neumann entropy\cite{brown2021computing,brown2024device}, and multi-basis key generation protocols\cite{schwonnek2021device}.

Advantage distillation (AD)\cite{tan2020advantage,liu2023mode,zhu2023reference,du2025advantage,stasiuk2025quantum}, a post-processing technique rooted in classical information theory, can effectively enhance the correlation of data among legitimate users and significantly improve the protocol’s robustness in noisy environments. This technique has already demonstrated excellent performance optimization in two-party QKD such as device-independent quantum key distribution (DI-QKD)\cite{tan2020advantage,stasiuk2025quantum}, mode-pairing quantum key distribution (MP-QKD)\cite{liu2023mode}, reference-frame-independent quantum key distribution (RFI-QKD)\cite{zhu2023reference} and phase-matching quantum key distribution (PM-QKD)\cite{wang2022phase}. Extending this technical framework to more complex multi-party quantum cryptographic protocols is a vital research direction for enhancing their practicality. However, the application of existing advantage distillation schemes in multi-party scenarios requires further exploration. Research on its extension to three-party DI-QSS protocols remains relatively limited. Meanwhile, although recent studies have begun to explore the application of AD in multi-party Quantum Conference Key Agreement (QCKA), demonstrating its potential in asymmetric noise networks\cite{krawec2025quantum}. Constructing an AD framework suitable for different multi-party interactive scenarios, such as three-party quantum secret sharing, and systematically evaluating its potential for enhancing key performance metrics remains a valuable subject for in-depth investigation.

A central contribution of this work is that the successful generalization and adaptation of the advantage distillation technical framework, originally developed mainly for two-party protocols, extends to the more complex three-party DI-QSS scenario. Compared to their two-party counterparts, three-party protocols involve more intricate interaction logic and security models and are more sensitive to noise, making the direct application of two-party solutions problematically. This work constructs an advantage distillation scheme suitable for the three-party collaborative work of Alice, Bob and Charlie by designing specific tripartite data interaction and consistency verification procedures. The systematic integration of this technique into the DI-QSS protocol\cite{zhang2024device} aims to decrease the original protocol's strong reliance on high detection efficiency and low QBER. Research demonstrates that this technique can effectively enhance the overall performance of the protocol. 

Specifically, through theoretical derivation and numerical simulation, it is verified that, when combined with advantage distillation, the protocol achieves significant optimization in core metrics such as noise tolerance, global detection efficiency threshold, and maximum secure communication distance. For example, when the entanglement fidelity $F$=0.98 and grobal detection efficiency $\eta=0.98$, the noise tolerance of basic protocol can be increased from 10.17$\%$ to 28.49$\%$ and the global detection efficiency threshold is lowered, enabling the protocol to generate secure information even with lower detection efficiency. When it combined with specific channel models, the maximum secure communication distance can be significantly extended from 0.16 km to 1.85 km. These optimizations provide a new technical pathway for the practical deployment of DI-QSS in practical noisy environments.

The subsequent structure of this paper is arranged as follows: Chapter~\ref{two} we elaborate on the fundamental principles of the DI-QSS protocol and the operational framework of the tripartite AD technique. Chapter~\ref{three} we detail the derivation of the theoretical formulas for the QBER and secret sharing rate after integrating the AD technique with the basic DI-QSS protocol and its three active improvement strategies (noise pre-processing, post-selection and their combined strategy). Chapter~\ref{four} we present a comprehensive comparative analysis, through systematic numerical simulation, of the optimization effects conferred by the AD technique on the performance metrics of the various protocols. Finally, Chapter~\ref{five} we provide a summary and outlook for the entire paper.

\section{Preliminaries}\label{two}
This chapter details the operational principles of the Device-Independent Quantum Secret Sharing (DI-QSS) protocol\cite{zhang2024device}, followed by an introduction to the framework of the tripartite advantage distillation technique.

\subsection{Device-Independent Quantum Secret Sharing protocol}

\begin{figure}[htbp]
	\centering
	\includegraphics[width=0.9\columnwidth]{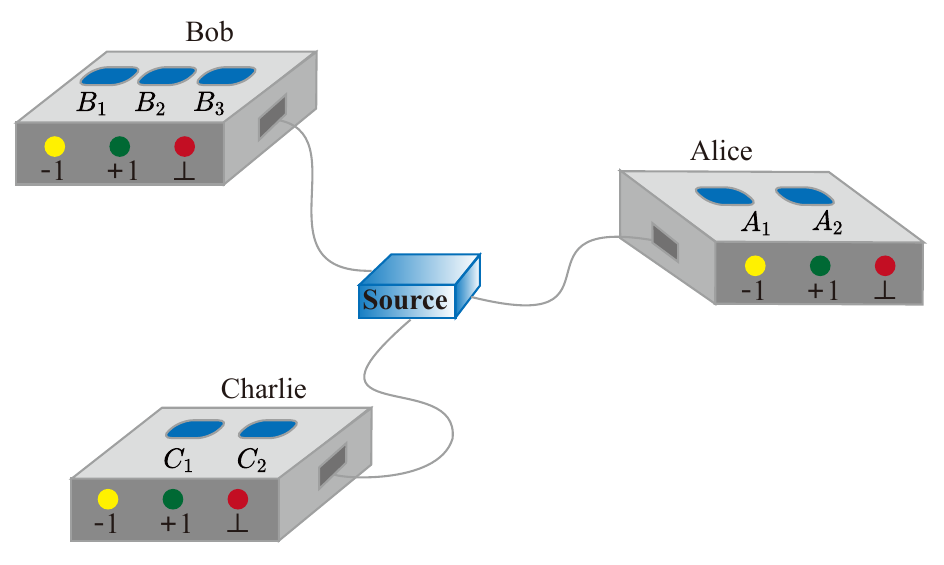}
	\caption{Schematic of the DI-QSS protocol. A central source prepares a large number of identical three-photon polarization-entangled GHZ states, which are then divided into three sequences, denoted as $E_{1}$, $E_{2}$ and $E_{3}$. The photons in these three sequences are distributed to three remote users, Alice, Bob and Charlie, respectively. Each user independently chooses a measurement basis and obtains a measurement outcome of -1, +1 or no-click event ($\bot$).}
	\label{fig.1}
\end{figure}
DI-QSS is an advanced quantum cryptographic protocol whose security does not rely on the internal working model of the experimental devices, but is based solely on the fundamental principles of quantum physics. The DI-QSS protocol\cite{zhang2024device} adopted in this work is implemented using a three-photon Greenberger-Horne-Zeilinger (GHZ) state. Specifically, the protocol employs a polarization GHZ state as the quantum resource, which takes the form:
\begin{equation}
	|\mathrm{GHZ}\rangle = \frac{1}{\sqrt{2}} (|\mathrm{HHH}\rangle + |\mathrm{VVV}\rangle)
	\label{eq:ghz_state} % 标签，用于引用
\end{equation}
where $ |\mathrm{H}\rangle $ and $ |\mathrm{V}\rangle $ denote the horizontal and vertical polarization states of a photon, respectively. The protocol involves three participants: Alice, Bob and Charlie.

Based on the schematic diagram of the protocol provided in Fig.~\ref{fig.1}, the core procedures and principles of the protocol are summarized as follows: 

Step 1. State Preparation and Distribution. A central entanglement source prepares a large number of three-photon GHZ states. The three photons from each GHZ state are sent to Alice, Bob and Charlie via quantum channels.

Step 2. Meamurement. Each user independently and randomly chooses a measurement basis to measure their received photon.
\begin{itemize}
	\item Alice's bases: $A_{1} = \sigma_{x}$, $A_{2} = \sigma_{y}$
	\item Bob's bases: $B_{1} = \sigma_{x}$, $B_{2} = (\sigma_{x} - \sigma_{y})/\sqrt{2}$, \\$B_{3} = (\sigma_{x} + \sigma_{y})/\sqrt{2}$
	\item Charlie's bases: $C_{1} = \sigma_{x}$, $C_{2} = -\sigma_{y}$
	\item The measurement results: $a_{i}, b_{j}, c_{k} \in \{+1, -1\}$
\end{itemize}

Step 3. Basis Announcement and Raw Secret Sifting. Parties announce their measurement basis publicly. Data is processed according to basis combinations. When Bob uses $B_{2}$ or $B_{3}$, all parties publish their results to estimate the Svetlichny polynomial  $S_{ABC}$. In this scenario, Eve’s eavesdropping cannot be detected by the users, rendering the photon transmission process insecure and necessitating the termination of the communication. For the combination $ \left\{ A_{1}, B_{1}, C_{1} \right\} $, results satisfying $ a = b \oplus c $ are kept as raw secret bits. Alice randomly discloses a subset of her raw secret bits, while Bob and Charlie reveal the corresponding raw secret bits to estimate the quantum bit error rate (QBER) $\delta$. The remaining undisclosed raw secret bits are retained by the parties. For the measurement basis combinations $ \left\{ A_{1}, B_{1}, C_{2} \right\} $, $ \left\{ A_{2}, B_{1}, C_{1} \right\} $ and $ \left\{ A_{2}, B_{1}, C_{2} \right\} $, the users have to discard their corresponding measurement results.

Step 4. Post-processing. The users iteratively perform the aforementioned steps until an adequate number of raw secret bits have been accumulated. Standard classical post-processing procedures including error correction and privacy amplification are performed to obtain a final secret for each party.

Step 5. Secret Reconstruction. Charlie announces his secret bits. Bob combines his own bits with Charlie's to reconstruct Alice's secret\cite{zhang2024device}.

Security Analysis: The protocol's security is based on the violation of the Svetlichny polynomial\cite{svetlichny1987distinguishing,mermin1990extreme}, a criterion for genuine tripartite nonlocality, defined as:
{
\allowdisplaybreaks[0]
\begin{multline}
	S_{ABC} = \langle a_{1} b_{2} c_{2} \rangle + \langle a_{1} b_{3} c_{1} \rangle + \langle a_{2} b_{2} c_{1} \rangle - \langle a_{2} b_{3} c_{2} \rangle \\
	+ \langle a_{2} b_{3} c_{1} \rangle + \langle a_{2} b_{2} c_{2} \rangle + \langle a_{1} b_{3} c_{2} \rangle - \langle a_{1} b_{2} c_{1} \rangle
\end{multline}
}
A value $ S_{ABC} > 4 $ certifies security. In practice, this is equivalent to a simplified CHSH polynomial\cite{bancal2011detecting}
\begin{equation}
	S = 2\sqrt{2} F \eta^{3}
\end{equation}
where $F$ is the GHZ state fidelity and $\eta$ is the global detection efficiency.

\subsection{Tripartite Advantage Distillation}
\begin{figure}[htbp]
	\centering
	\includegraphics[width=0.9\columnwidth]{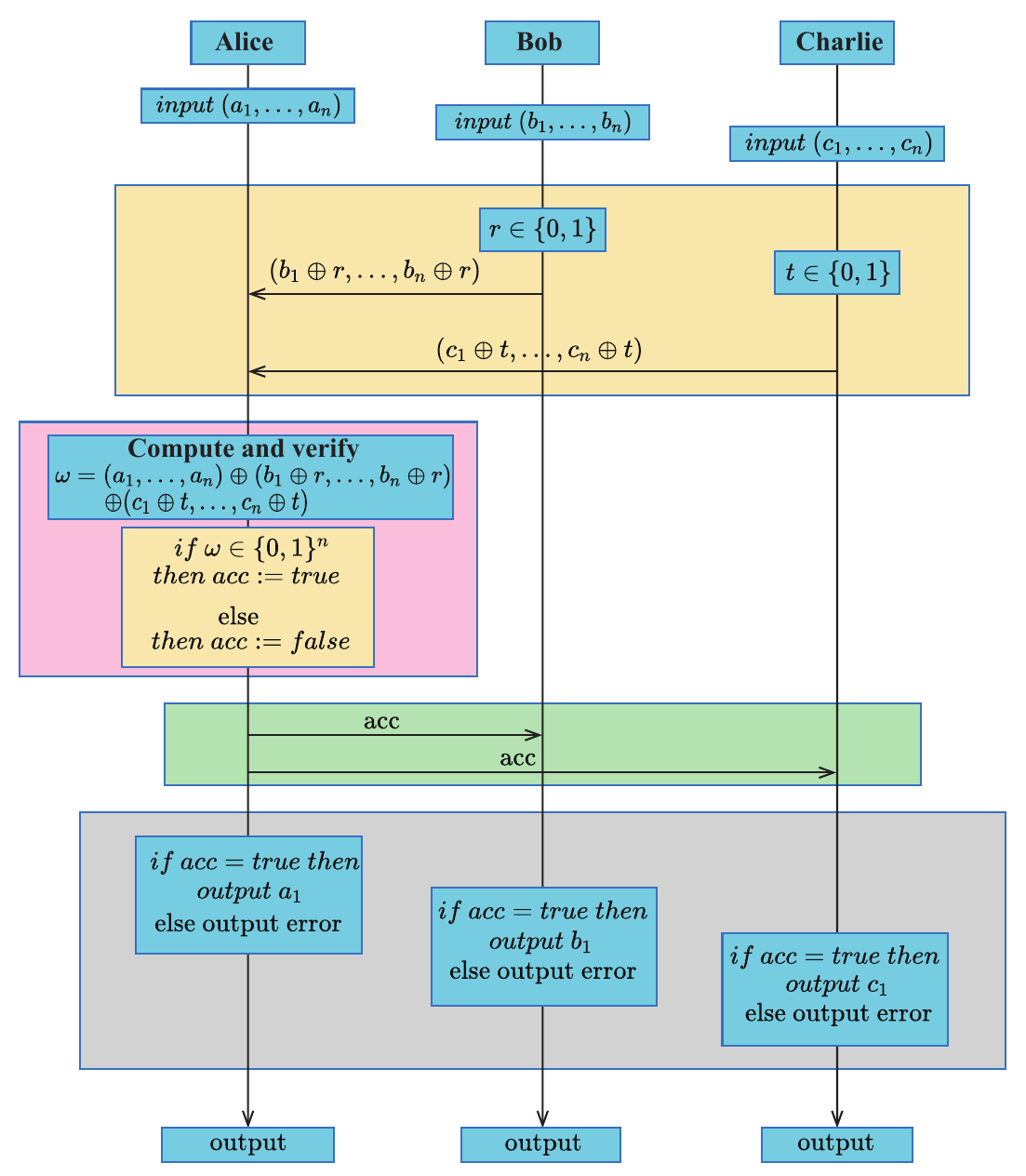}
	\caption{Flowchart of the tripartite advantage distillation procedure. This figure illustrates the post-processing procedure of the advantage distillation technique in a tripartite DI-QSS protocol. }
	\label{fig.2}
\end{figure}
Assume Alice, Bob and Charlie hold raw secret bit sequences $\left( a_{1}, \ldots, a_{m} \right)$, $\left( b_{1}, \ldots, b_{m} \right)$ and $\left( c_{1}, \ldots, c_{m} \right)$ respectively. Ideally, the relationship $ a_i = b_i \oplus c_i $ should hold for each bit position \textit{i}, but errors may occur in practice due to noise, collective attacks and coherent attacks which leads to the QBER $\delta$. The advantage distillation process is as follows: The detailed procedure of the protocol is illustrated in Fig.~\ref{fig.2}. Users divide their sequences into blocks of length \textit{n}. For each block, Bob and Charlie generate random bits $ r, t \in \{0, 1\} $
respectively, and send $ \left(b_{1},\ldots, b_{n}\right)\oplus(r,\ldots, r)\text{ and }\left(c_{1},\ldots, c_{n}\right)\oplus(t,\ldots, t) $ to Alice over a classical channel. Alice then computes:
\begin{equation}
	\begin{aligned}
		 (a_1, \ldots, a_n) &\oplus \bigl[(b_1, \ldots, b_n) \oplus (r, \ldots, r)\bigr] \\
		 \quad &\oplus \bigl[(c_1, \ldots, c_n) \oplus (t, \ldots, t)\bigr]
	\end{aligned}
\end{equation}
If the result is $(0,\dots,0)$ or $(1,\dots,1)$, the first bit of the block from each user $( a_{1}, b_{1}, c_{1} )$ is retained as part of the refined secret. Otherwise, the entire block is discarded.

This process acts as a filter, retaining data only when the error pattern within a block is consistent, thereby enhancing the correlation of the remaining secret bits. Advantage distillation can effectively reduce the QBER. The refined QBER after advantage distillation is derived by calculating the post-processing error probability under the specific rule of using a block length of 2. This choice of block length represents a deliberate trade-off between performance and efficiency. While longer blocks (eg. 3, 4, or 5 bits) may offer stronger error-filtering capability, they also lead to a sharp decline in the data retention rate. The block-length-2 scheme, however, can significantly enhance core performance metrics such as the protocol's noise tolerance and lower the global detection efficiency threshold, while maintaining a relatively high data utilization rate. This achieves a more favorable balance between optimization effectiveness and resource consumption. Although a single round of extraction incurs data loss, it substantially increases the noise tolerance, providing stronger practicality for applications such as quantum secret sharing.

\section{Application of Advantage Distillation in Device-Independent Quantum Secret Sharing}\label{three}
In this section, for the basic DI-QSS protocol and its three active improvement strategies (noise pre-processing, post-selection, and noise pre-processing based post-selection), based on the working principle of tripartite advantage distillation, we establish the relationship between the pre- and post-distillation quantum bit error rates (QBER) and calculate the latter from the former. Furthermore, the theoretical formula for the secret sharing rate after applying the advantage distillation step is provided.

\subsection{Integration of the Basic DI-QSS Protocol with Advantage Distillation}
In the DI-QSS protocol, we only require Eve to obey the laws of quantum physics but place no restrictions on its capabilities. Users rely solely on the security certified by tripartite nonlocality to ensure the security of the secret. Under the white noise model, considering both decoherence and photon loss, the QBER $\delta$  for the basic DI-QSS protocol is\cite{zhang2024device}:
\begin{equation} 
	\delta = 1 - \frac{1}{2} \eta^{3} (1 + F) 
	\label{eq.6}
\end{equation}

Under collective attacks, the secret sharing rate is estimated using the Devetak-Winter bound\cite{devetak2005distillation,augusiak2009multipartite}. The final tripartite secret sharing rate $r$ is:
\begin{equation}
	\begin{aligned}
		r &= H(A_{1} \mid E) - H(A_{1} \mid B_{1}, C_{1}) \\
		&\geq 1 - h\left(\frac{\sqrt{S^{2}/4 - 1}}{2} + \frac{1}{2}\right) - h(\delta)
	\end{aligned}
\end{equation}
Here, the theoretical value of the CHSH polynomial $S$ is $S = 2\sqrt{2} F \eta^{3}$.

According to the rule of advantage distillation, instances where any legitimate party fails to detect a photon are filtered out, retaining only cases where all three parties successfully detect photons. After the tripartite AD with a block length of 2 bits, the QBER $\delta_{ad}$ is given by Eq.~\eqref{eq.5}, and the detailed proof is provided in Appendix~\ref{A}.
\begin{equation}\label{eq.5}
	\delta_{ad} = \frac{(1-F)^{2}}{(1+F)^{2}+(1-F)^{2}}
\end{equation}
The final tripartite secret sharing rate $r_{ad}$ is then:
\begin{equation}
	r_{ad} \geq 1 - h \left( \frac{\sqrt{S^{2}/4 - 1}}{2} + \frac{1}{2} \right) - h \left( \delta_{a d} \right) 
\end{equation}

\subsection{Integration of DI-QSS with Noise Pre-processing and Advantage Distillation}
The noise pre-processing strategy enhances noise resistance by intentionally adding a certain amount of noise to the initial measurement data. Specifically, Alice performs the pre-processing operation in Step 3 of the DI-QSS protocol. When the measurement basis settings for all three parties are $ \left\{ A_{1} = B_{1} = C_{1} = \sigma_{x} \right\} $, Alice flips her measurement outcome (changing +1 to -1, and -1 to +1) with a probability $q$. Such a noise pre-processing method in this protocol is similar to that adopted in DI-QKD\cite{ho2020noisy,woodhead2021device}. The additional noise degrades the correlation between the secret bits of Alice and Bob, as well as the correlation between the secret bits of Alice and Eve. The secret sharing rate depends on the difference between these two correlations. Therefore, the overall effect may potentially increase the secret sharing rate. After all photons have been measured and the flip operations are completed, Alice announces the flip probability $q$ during the error correction process.

After noise pre-processing, the overall QBER $\delta_{q}$ can be readily obtained\cite{zhang2024device}:
\begin{equation}\label{eq.9}
	\delta_{q} = q(1-\delta) + (1-q)\delta = q + (1-2q)\delta 
\end{equation}

Based on the derivation from the DI-QKD protocol employing the noise pre-processing strategy\cite{ho2020noisy,woodhead2021device,sekatski2021device}, the lower bound on the secret sharing rate after noise pre-processing with respect to Eve can be estimated as
\begin{equation}
	H\left(A_{1} \mid E\right)_{q} \geqslant g(S, q) 
\end{equation}
where
\begin{equation}
	\begin{split}
		g(S,q) &= 1 - h \left( \frac{\sqrt{S^{2}/4 - 1}}{2} + \frac{1}{2} \right) \\
		&\quad + h \left( \frac{\sqrt{(1-2q)^{2} + 4q(1-q)\left({S^{2}/4} - 1\right)}}{2} + \frac{1}{2} \right)
	\end{split}
\end{equation}
The final tripartite secret sharing rate $r_{q}$ then becomes\cite{zhang2024device}
\begin{equation}
	r_{q} \geq g(S, q) - h\left(\delta_{q}\right)
	\label{eq:rq_inequality_2col}
\end{equation}

With the incorporation of the advantage distillation technique and according to Eq.~\eqref{eq.9}, the QBER $\delta_{q-ad}$ for the DI-QSS protocol under the noise preprocessing strategy is given by:
\begin{equation}
	\delta_{q-ad} = q + (1 - 2q) \delta_{ad} 
\end{equation}
The final tripartite secret sharing rate $r_{q-ad}$ then becomes
\begin{equation}
	r_{q-ad} \geq g\left(S, q\right) - h\left(\delta_{q-a d}\right)
\end{equation}

\subsection{Integration of DI-QSS with Post-Selection Strategy and Advantage Distillation}
All three users employ a post-selection strategy to enhance the photon loss tolerance of DI-QSS. This post-selection method accounts for all detector response and non-response events. In the second step of the protocol, when the measurement basis combination for all three parties is $ \left\{A_{1},B_{1},C_{1}\right\} $, each user replaces the previous three-valued strategy with a deterministic binary strategy. Specifically, besides the deterministic outcomes $ +1\left(\left|+x\right\rangle=\frac{1}{\sqrt{2}}\left(|0\rangle+|1\rangle\right)\right)\text{ and }-1\left(\left|-x\right\rangle=\frac{1}{\sqrt{2}}\left(|0\rangle-|1\rangle\right)\right)$, users also re-map the detector's non-response outcome from the symbol ``$\perp$'' to ``+1''. The post-selection strategy maps the detector response status as: $``+"\rightarrow``+1", ``-"\rightarrow``-1", ``\perp"\rightarrow``+1"$ The corresponding relationships are specifically:
\begin{align*}
	\{+ + \perp\},\{+ \perp +\},\{\perp + +\},\{+ \perp\perp\},\\\{\perp\perp +\},\{\perp + \perp\},\{\perp\perp\perp\} &\to \{+ + +\} \\
	\{\perp + -\},\{+ \perp -\},\{\perp\perp -\} &\to \{+ + -\} \\
	\{+ - \perp\},\{\perp - +\},\{\perp - \perp\} &\to \{+ - +\} \\
	\{- + \perp\},\{- \perp +\},\{- \perp\perp\} &\to \{- + +\} \\
	\{\perp - -\} &\to \{+ - -\} \\
	\{- - \perp\} &\to \{- - +\} \\
	\{- \perp -\} &\to \{- + -\}
\end{align*}

After applying the post-selection strategy, the probability of the measurement outcomes will change to\cite{zhang2024device}:
\begin{align*}
	P_{p}(+ + +) &= P(+ + +) + P(+ + \perp) + P(+ \perp +) \notag \\
	&\quad + P(\perp + +) + P(+ \perp \perp) + P(\perp + \perp) \notag \\
	&\quad + P(\perp \perp +) + P(\perp \perp \perp), \\[6pt]
	P_{p}(+ - -) &= P(+ - -) + P(\perp - -), \\[6pt]
	P_{p}(- + -) &= P(- + -) + P(- \perp -), \\[6pt]
	P_{p}(- - +) &= P(- - +) + P(- - \perp), \\[6pt]
	P_{p}(+ + -) &= P(+ + -) + P(+ \perp -) + P(\perp + -) \notag \\
	&\quad + P(\perp \perp -), \\[6pt]
	P_{p}(+ - +) &= P(+ - +) + P(+ - \perp) + P(\perp - +) \notag \\
	&\quad + P(\perp - \perp), \\[6pt]
	P_{p}(- + +) &= P(- + +) + P(- + \perp) + P(- \perp +) \notag \\
	&\quad + P(- \perp \perp), \\[6pt]
	P_{p}(- - -) &= P(- - -).
\end{align*}
And the tripartite quantum state evolves to
\begin{equation}
	\begin{aligned}
		\rho_{ABC}^{\prime} &= \eta^{3}\left[ F\ket{\phi_1}\bra{\phi_1} + \frac{1-F}{8} I \right] \\
		&\quad + \frac{1}{2}\eta^{2}\bar{\eta} \left( \ket{+HH}\bra{+HH} + \ket{+VV}\bra{+VV} \right) \\
		&\quad + \frac{1}{2}\eta^{2}\bar{\eta} \left( \ket{H+H}\bra{H+H} + \ket{V+V}\bra{V+V} \right) \\
		&\quad + \frac{1}{2}\eta^{2}\bar{\eta} \left( \ket{HH+}\bra{HH+} + \ket{VV+}\bra{VV+} \right) \\
		&\quad + \frac{1}{2}\eta\bar{\eta}^{2} \left( \ket{H++}\bra{H++} + \ket{V++}\bra{V++} \right) \\
		&\quad + \frac{1}{2}\eta\bar{\eta}^{2} \left( \ket{+H+}\bra{+H+} + \ket{+V+}\bra{+V+} \right) \\
		&\quad + \frac{1}{2}\eta\bar{\eta}^{2} \left( \ket{++H}\bra{++H} + \ket{++V}\bra{++V} \right) \\
		&\quad + \bar{\eta}^{3} \ket{+++}\bra{+++}
	\end{aligned}
\end{equation}

When the measurement outcomes $a, b, c$ are 000, 011, 101 or 110, the results of the three parties satisfy $a \oplus b \oplus c = 0$. And when the measurement outcomes $a, b, c$ are 001, 010, 100 or 111, the results of the three parties do not satisfy $a \oplus b \oplus c = 0$, which constitutes an error. Therefore, under the post-selection strategy, the QBER $\delta_{p}$  is 
\begin{align}
	\delta_{p} = \frac{1 - F}{2} \eta^{3} - \frac{3}{2} \eta^{2} + \frac{3}{2} \eta
	\label{eq.17}
\end{align}
The final tripartite secret sharing rate $r_{p}$ is
\begin{equation}
	\begin{aligned}
		r_{p} \geq 1 - h\left( \frac{\sqrt{S_{p}^{2}/4 - 1}}{2} + \frac{1}{2} \right) - h\left(\delta_{p} \right)
	\end{aligned}
	\label{eq:rp_formula}
\end{equation}
Here, adopting the post-selection strategy alters the CHSH polynomial, so $S$ will change accordingly and becomes\cite{zhang2024device}:
\begin{equation}
	S_{p} = 2\sqrt{2} F\eta^{3} + 2\bar{\eta}^{3}
\end{equation}

The value of this expression is greater than the original value of $S$. Substituting it into $r_{p}$, we find that as the value of the CHSH polynomial increases, the secret secrecy rate against Eve can be effectively improved. After incorporating the advantage distillation technique, the QBER $\delta_{p-ad}$ for the tripartite execution with a block length of 2 is calculated as:
\begin{equation}
	\delta_{p-\mathrm{ad}}=
	\frac{\left(1-F\eta^{3}-\bar{\eta}^{3}\right)^{2}}
	{\left(1+F\eta^{3}+\bar{\eta}^{3}\right)^{2}
		+\left(1-F\eta^{3}-\bar{\eta}^{3}\right)^{2}}
	\label{eq.20}
\end{equation}
The proof of Eq.~\eqref{eq.20} is analogous to that in Appendix~\ref{A}. And the final tripartite secret sharing rate $r_{p-ad}$ becomes:
\begin{equation}
	\begin{aligned}
		r_{p-\mathrm{ad}} &\geq  1 - h\left( \frac{\sqrt{S_{p}^{2}/4 - 1}}{2} + \frac{1}{2} \right) - h\left( \delta_{p-ad} \right)
	\end{aligned}
	\label{eq:main_inequality_alt}
\end{equation}

\subsection{Integration of DI-QSS with the Post-Selection Strategy Combined with Noise Pre-processing and Advantage Distillation}
This strategy achieves higher noise tolerance by integrating post-selection with noise preprocessing. After applying this active improvement strategy, we can derive the total QBER $\delta_{qp}$ for the DI-QSS protocol as
\begin{equation}
	\begin{aligned}
		\delta_{qp} &= q + (1 - 2q)\delta_{p} 
	\end{aligned}
	\label{eq:delta_qp}
\end{equation}
Since noise preprocessing does not affect the CHSH polynomial, we have $S_{qp}=S_{p}$. Therefore, the tripartite secret sharing rate $r_{qp}$ is\cite{zhang2024device}
\begin{equation}
	\begin{aligned}
		r_{qp} \geq g( S_{qp}, q ) - ( \delta_{qp} )
	\end{aligned}
\end{equation}

After applying the advantage distillation technique, the error rate for noise preprocessing under advantage distillation becomes:
\begin{equation}
	\delta_{q p\text{-}ad} = q + (1 - 2q)\delta_{p\text{-}ad}
\end{equation}
Finally, the secret sharing rate $r_{qp-ad}$ for the post-selection strategy based on noise preprocessing after advantage distillation can be obtained as:
\begin{equation}
	\begin{aligned}
		r_{qp-ad} \geq (S_{qp}, q) - h( \delta_{qp-ad} )
	\end{aligned}
\end{equation}

\section{Numerical Simulation and Performance Analysis}\label{four}
To evaluate the performance of the basic DI-QSS protocol and the three active improvement strategies combined with advantage distillation in practical communication scenarios, we conducted systematic simulation experiments and performed a comparative analysis of the results. In the simulations, we set different values for the noise pre-processing flip probability $q$, and analyzed the optimization effects of the advantage distillation technique on the secret sharing rate, the global detection efficiency threshold, noise tolerance, and the maximum secure communication distance under various strategies.

\subsection{Advantage Distillation and the Basic DI-QSS Protocol}

\begin{figure}[htbp]
	\centering
	\includegraphics[width=0.9\columnwidth]{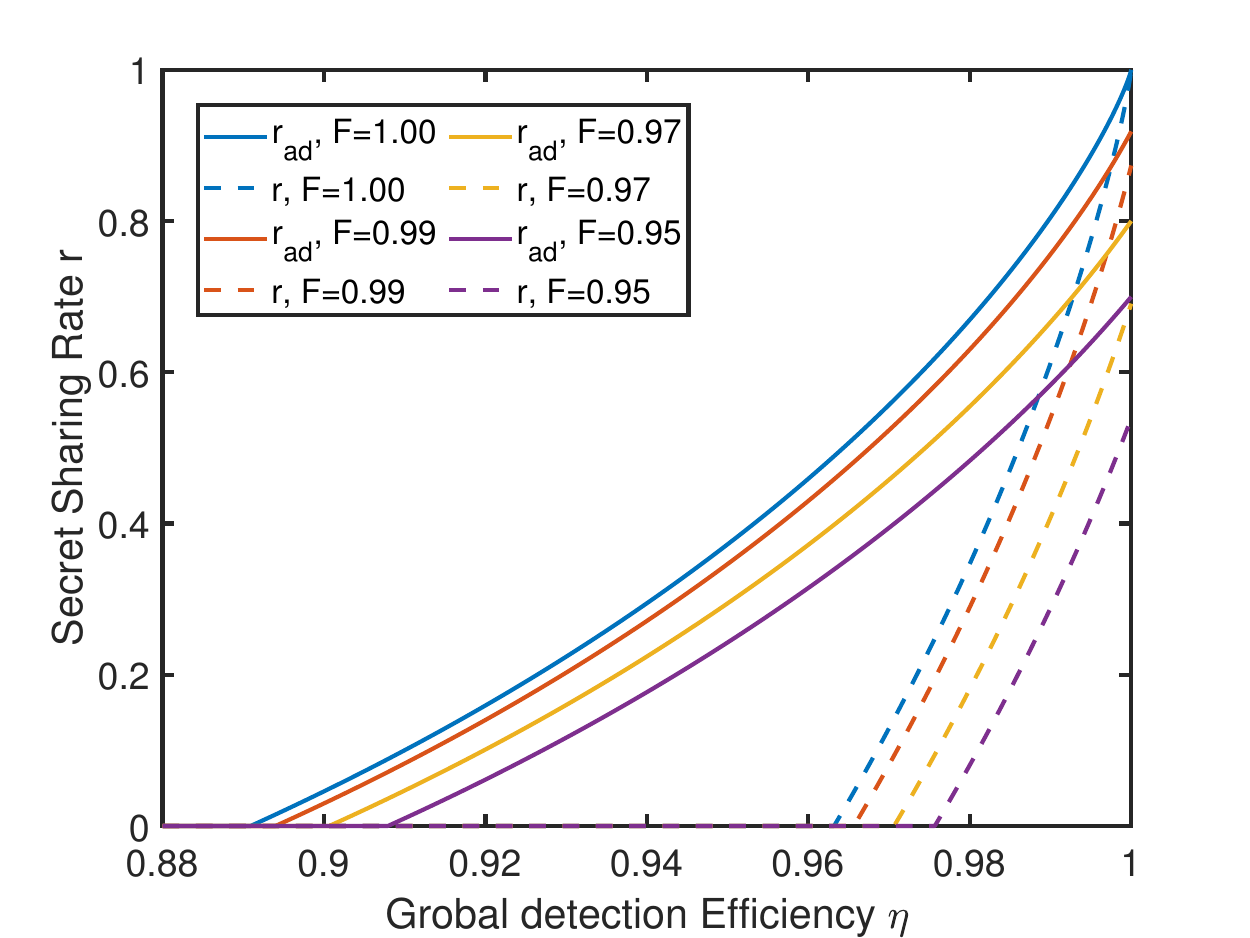}
	\caption{Comparison of the secret sharing rates ($r$) before and after advantage distillation under different fidelities. Dashed and solid lines respectively represent the protocol performance before and after advantage distillation under various fidelities. It can be intuitively observed that after applying advantage distillation, the $\eta-r$ curves under all fidelities shift upward significantly}
	\label{fig.4}
\end{figure}

By applying AD to the DI-QSS protocol, as shown in Fig.~\ref{fig.4}, the figure illustrates the relationship between the secret sharing rate $r$ and the global detection efficiency $\eta$ under different fidelities. Before AD, when the entanglement fidelities are $F=1$, $F=0.99$, $F=0.97$ and $F=0.95$, the corresponding global detection efficiency thresholds are $\eta=0.963$, $\eta=0.966$, $\eta=0.971$ and $\eta=0.976$, respectively. After AD, the global detection efficiency thresholds become $\eta=0.891$, $\eta=0.894$, $\eta=0.901$ and $\eta=0.908$, respectively. Moreover, under the same global detection efficiency, a higher fidelity results in a higher secret sharing rate. Additionally, as the global detection efficiency increases, the growth trend of the secret sharing rate becomes more pronounced. The results visually demonstrate that the requirement for global detection efficiency to generate a secure secret is reduced. This relaxes the demands on the environment and noise tolerance, achieving favorable performance enhancement.
\begin{figure}[htbp]
	\centering
	\includegraphics[width=0.8\columnwidth]{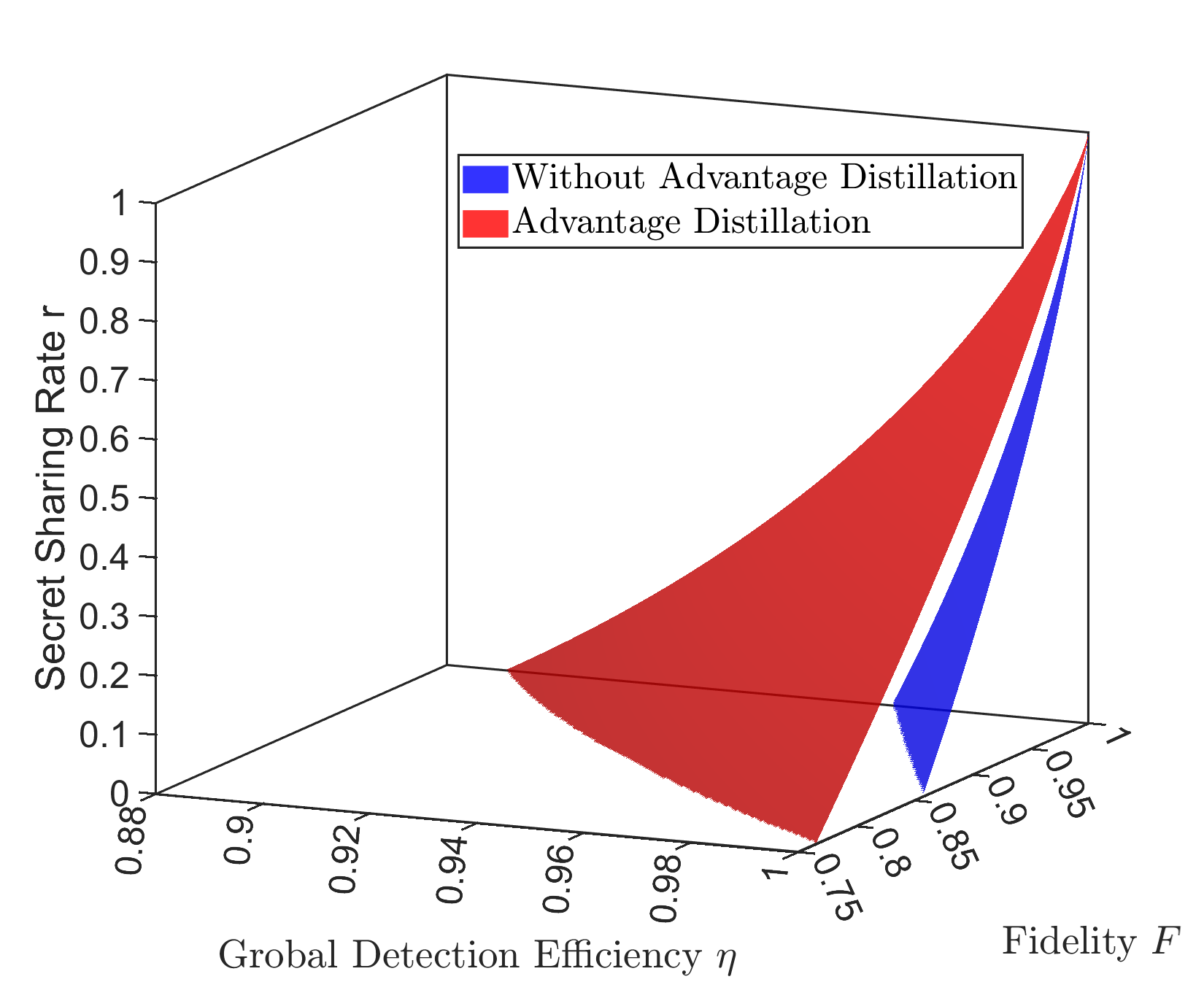}
	\caption{Three-dimensional plot of the secret sharing rate ($r$) versus the global detection efficiency ($\eta$) and fidelity ($F$) before and after applying advantage distillation. The blue surface (without advantage distillation) lies entirely below the red surface (with advantage distillation), indicating that the advantage distillation technique significantly enhances the secret sharing rate for any given combination of $\eta$ and $F$. More importantly, the red surface maintains a positive secret sharing rate ($r > 0$) even within regions of lower global detection rate and lower fidelity.}
	\label{fig.5}
\end{figure}

In the three-dimensional model, with the global detection efficiency $\eta$ as the x-axis, the entanglement fidelity $F$ as the y-axis, and the secret sharing rate $r$ as the z-axis, as shown in Fig.~\ref{fig.5}, we can plot the relationship among these three parameters before and after advantage distillation. From this plot, we can clearly observe that after advantage distillation, the requirement for global detection efficiency is lower. Specifically, when $0.89 < \eta < 0.963$, the secret sharing rate is zero before advantage distillation, whereas after advantage distillation, secure secrets can be generated within this entire interval. Furthermore, when $0.963 \leq \eta < 1$, the secret sharing rate after advantage distillation is significantly higher than that before advantage distillation.

\subsection{Advantage Distillation and DI-QSS with the Noise Pre-processing Strategy}
After applying noise pre-processing to the DI-QSS protocol, the noise tolerance of the protocol can be effectively enhanced. However, the excessively high signal flip probability introduced during the noise pre-processing process increases the error rate in signal transmission, thereby leading to a reduction in the secret sharing rate. Therefore, when deciding whether to adopt noise pre-processing, it is essential to comprehensively evaluate its ability to maximize the generation efficiency of secure secrets.

\begin{figure}[htbp]
	\centering
	\includegraphics[width=0.9\columnwidth]{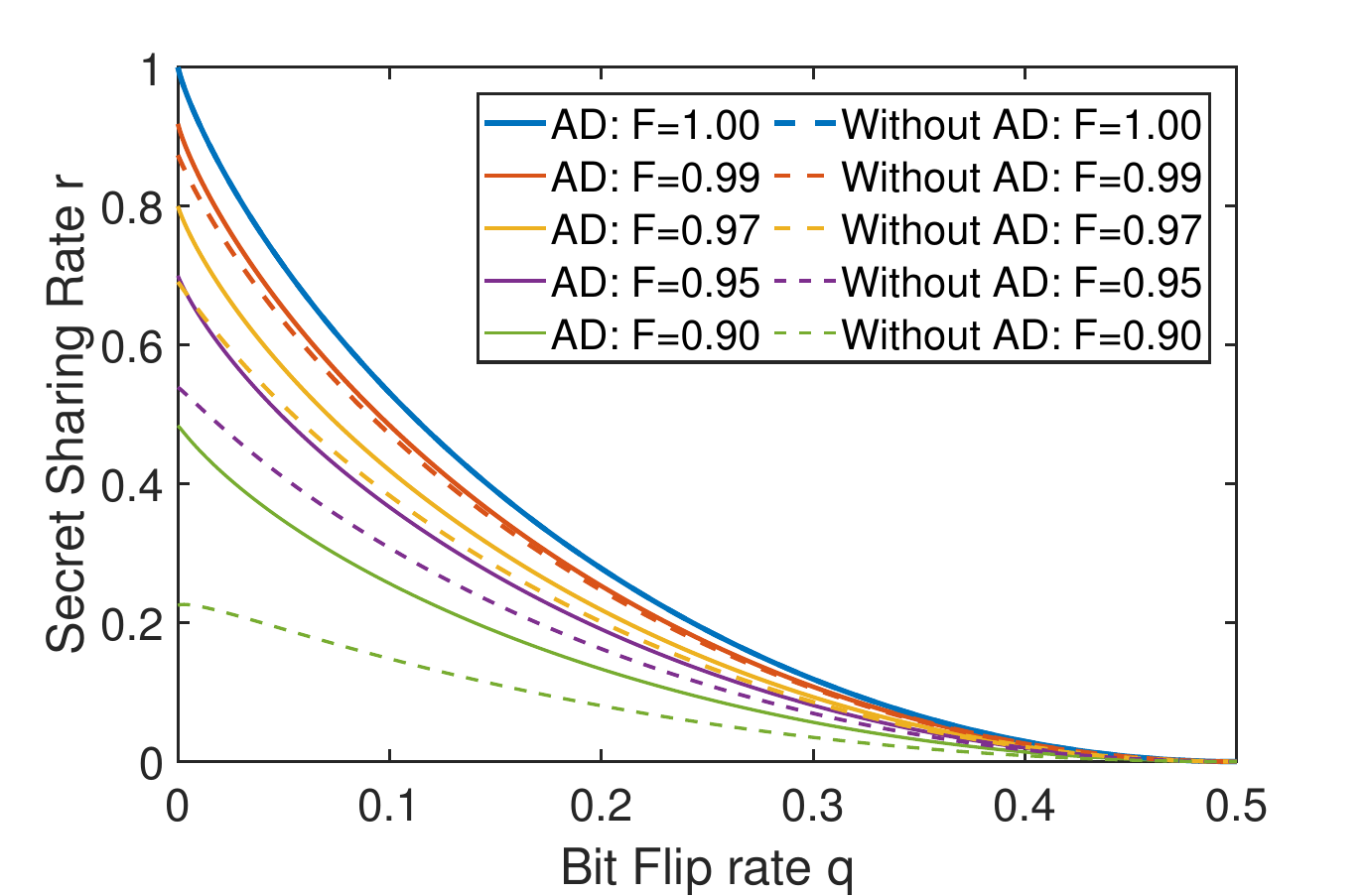}
	\caption{Dependence of the secret sharing rate $r$ on the bit-flip rate $q$ of noise preprocessing. The figure illustrates the variation of the secret sharing rate $r$ with the bit-flip rate $q$ for the DI-QSS protocol under different entanglement fidelities $F$, with (solid lines) and without (dashed lines) advantage distillation. It can be observed that the advantage distillation technique significantly improves the secret sharing rate of the protocol under the noise preprocessing strategy. Moreover, it is notably observed that the lower the fidelity, the more pronounced the improvement in the secret sharing rate after applying advantage distillation.}
	\label{fig.7}
\end{figure}
We now take five values of entanglement fidelity 1, 0.99, 0.97, 0.95 and 0.9 to compare the effect of the same signal flip probability on the secret sharing rate before and after advantage distillation. As shown in the Fig.~\ref{fig.7}, under different fidelity-based noise preprocessing strategies, the advantage distillation technique can significantly enhance the secret sharing rate of DI-QSS, particularly within the interval of $0 \leq q \leq 0.2$, where the effect after advantage distillation is notably pronounced. Taking $\eta$=1, it shows that a lower signal flip rate leads to a higher secret sharing rate. Under the same signal flip rate, a greater fidelity results in a higher secret sharing rate, with the threshold for the signal flip rate being 0.5 in all cases.

\begin{figure}[htbp]
	\centering
	\includegraphics[width=0.9\columnwidth]{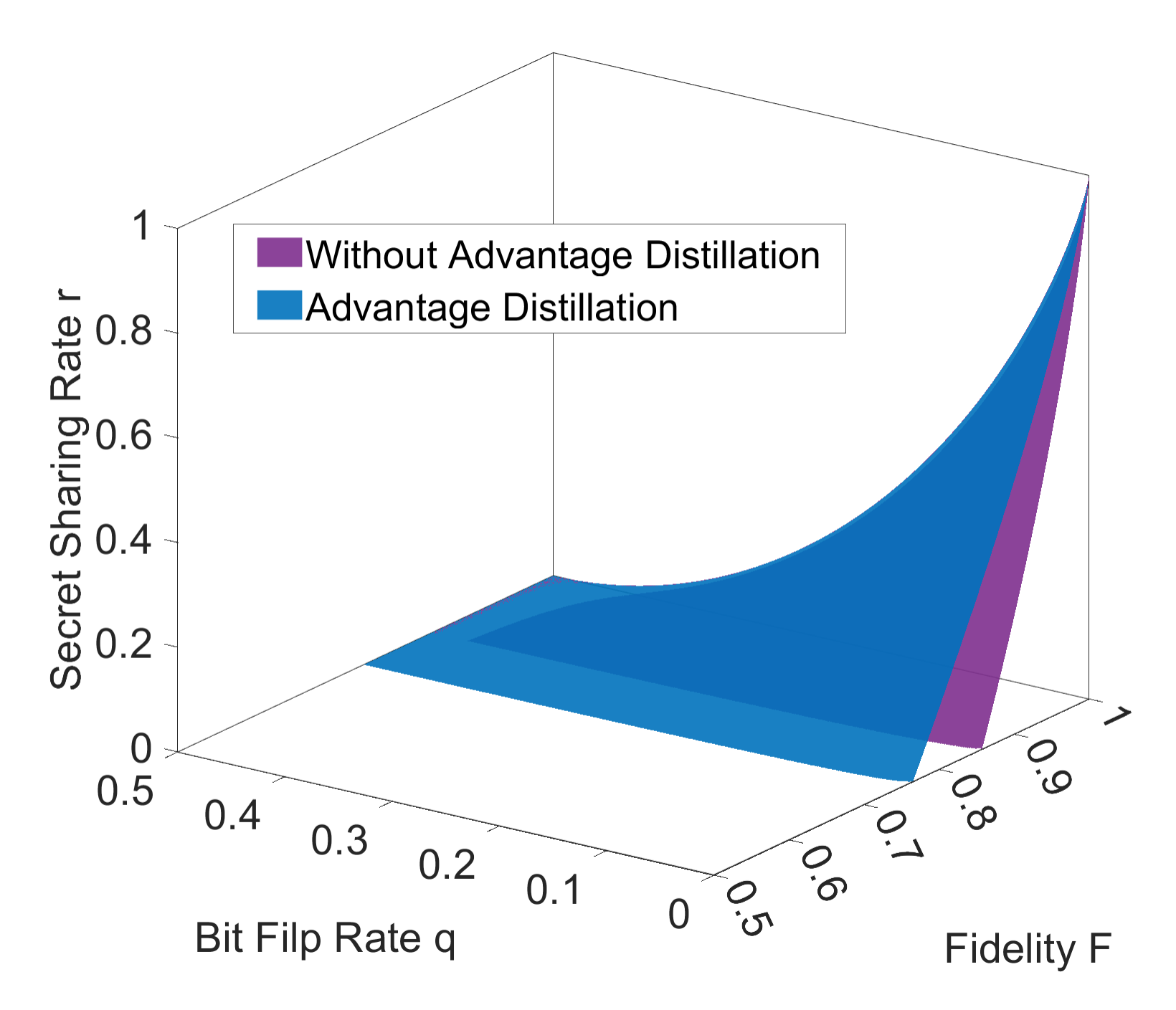}
	\caption{Three-dimensional surfaces showing the relationship between the secret sharing rate, bit‑flip rate, and fidelity. The blue surface represents the protocol performance with advantage distillation, while the purple surface corresponds to the original protocol performance. The plot demonstrates that for any given fidelity $F$, the advantage distillation technique significantly enhances the protocol’s ability to generate secure secrets under higher bit‑flip noise.}
	\label{fig.8}
\end{figure}

By taking the fidelity $F$ as one of the variables, a three-dimensional relationship diagram can be plotted among the fidelity $F$, qubit flip rate $q$, and secret sharing rate $r$ when the grobal detection efficiency $\eta=1$. From the Fig.~\ref{fig.8}, it can be observed that after advantage distillation, the secret sharing rate $r$ exhibits varying degrees of improvement across all scenarios.

\subsection{Advantage Distillation and DI-QSS with the Post-Selection Strategy}
\begin{figure}[htbp]
	\centering
	\includegraphics[width=0.9\columnwidth]{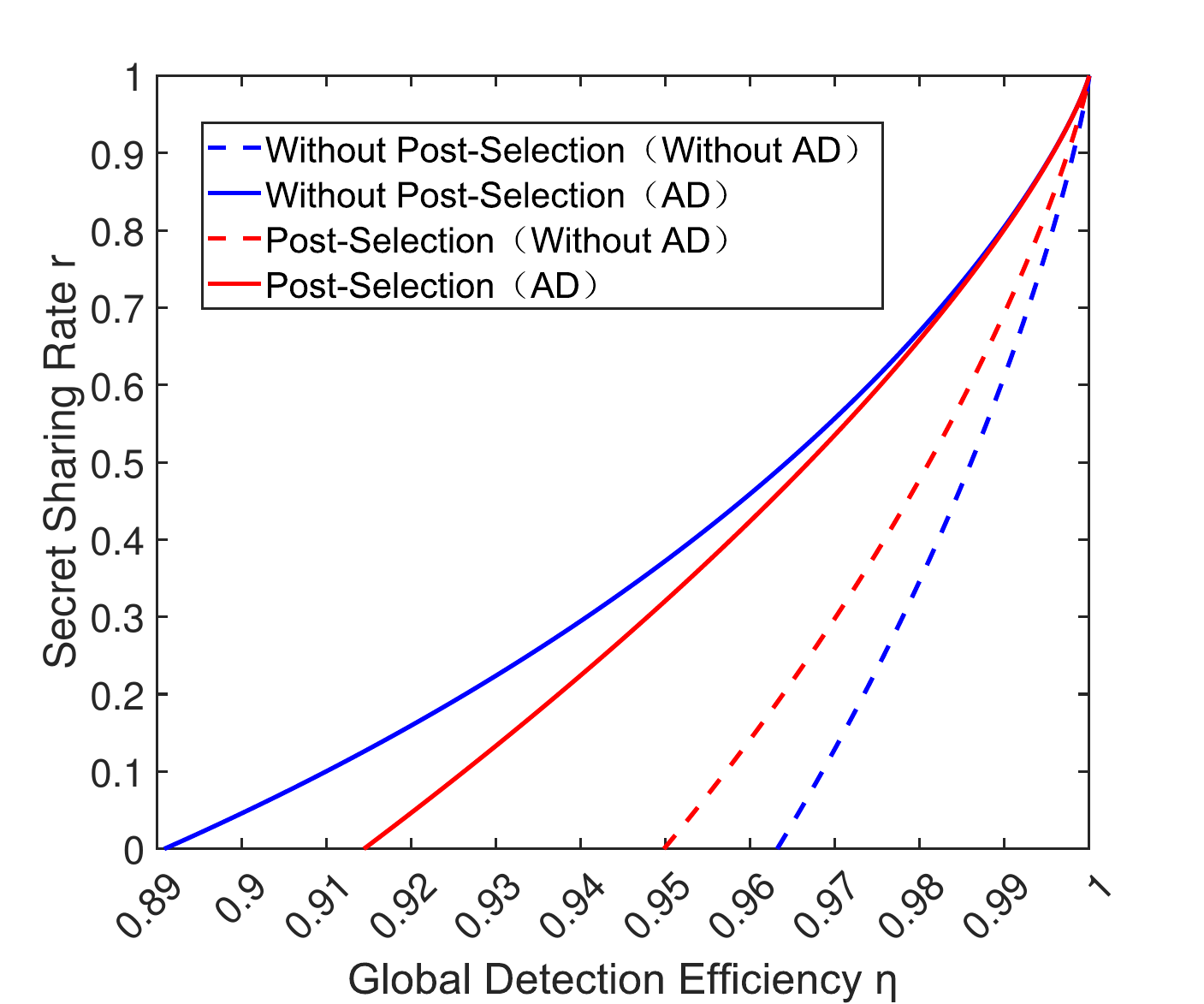}
	\caption{Curve of secret sharing rate versus global detection efficiency, pre- and post-advantage distillation. This figure systematically compares the variation of the secret sharing rate $r$ with the global detection efficiency $\eta$ under four combination conditions: with or without the post-selection strategy (red series or blue series curves) and with or without the advantage distillation technique (solid lines or dashed lines).}
	\label{fig.10}
\end{figure}

The DI-QSS protocol employing the post-selection strategy can reduce the global detection efficiency threshold\cite{zhang2024device}. After applying the advantage distillation technique, we analyzed that the DI-QSS protocol, regardless of whether the post-selection strategy is adopted, achieves a certain reduction in the global detection efficiency threshold. Also, when $F=1$, it can be observed from the Fig.~\ref{fig.10} that the global detection efficiency threshold of the protocol using the post-selection strategy decreases from 94.98$\%$ to 91.44$\%$, while the threshold of the protocol not using the post-selection strategy decreases from 96.32$\%$ to 89.09$\%$, representing reductions of 3.54$\%$ and 7.23$\%$, respectively. The effect of lowering the global detection efficiency threshold is relatively significant, which facilitates secure secret sharing even under lower detection efficiency conditions.

The optimization effect of advantage distillation depends on the gap in genuine quantum correlations between the users and the eavesdropper, the larger the gap, the better the optimization effect. In the DI-QSS protocol employing the post-selection strategy, classical random noise (denoted as $\bot$) is labeled as valid quantum signals (+1). This type of classical random error is independent of quantum entanglement and is entirely random for both the users and Eve. As a result, after the post-selection operation, a substantial amount of spurious correlation is mixed into the total correlation among the users, ultimately significantly narrowing the gap in genuine quantum correlations with Eve. This leads to a weaker advantage distillation effect compared to the basic DI-QSS protocol.

\begin{figure}[htbp]
	\centering
	\includegraphics[width=0.9\columnwidth]{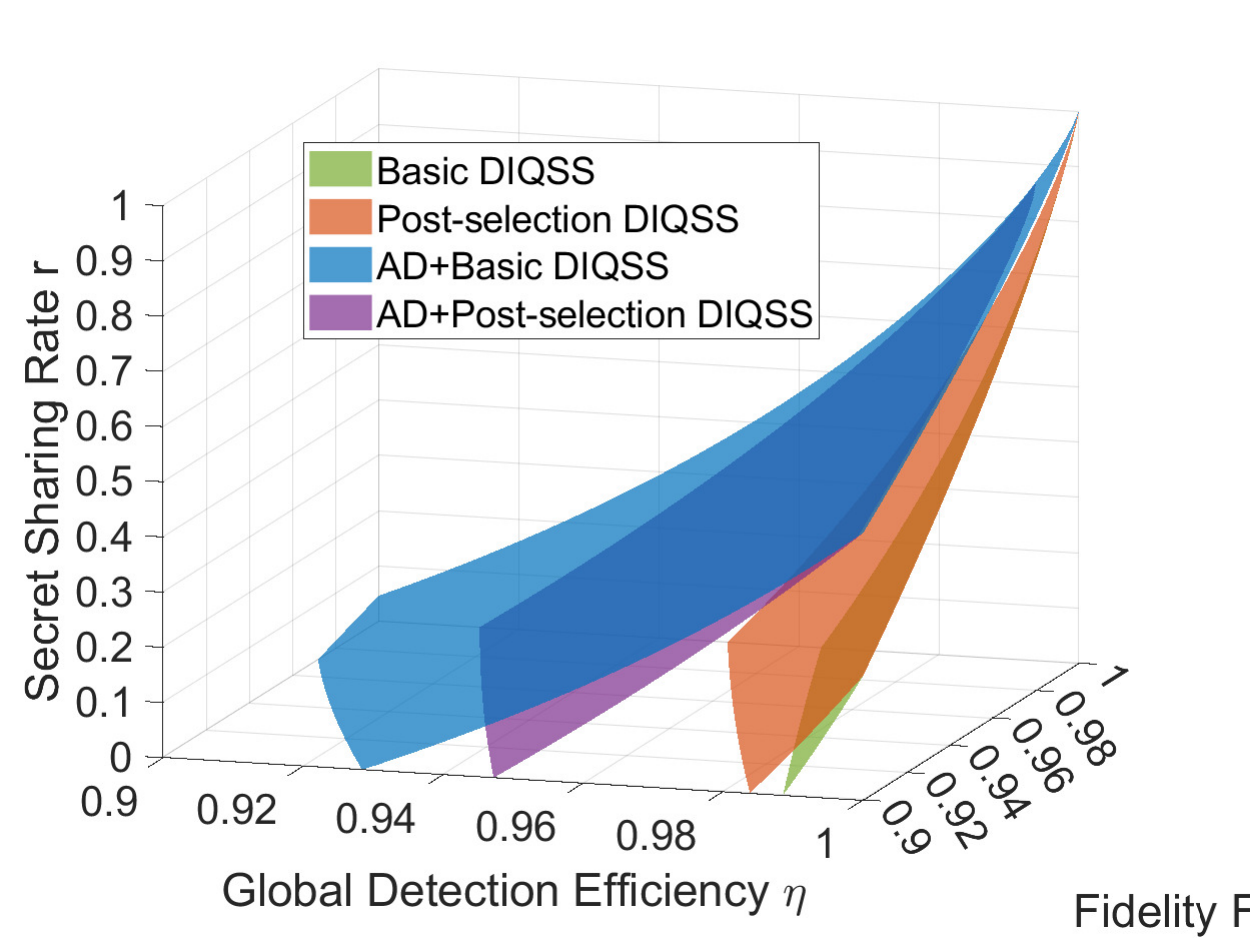}
	\caption{Three-dimensional surfaces of the secret sharing rate $r$ as a function of the global detection efficiency $\eta$ and the fidelity $F$. The figure demonstrates that advantage distillation can significantly enhance the secret sharing rate of the protocol over a wide range of $\eta$ and $F$.}
	\label{fig.11}
\end{figure}

As shown in the three-dimensional space of Fig.~\ref{fig.11}, through numerical analysis, when $0.9 < F < 1$, the advantage distillation technique can reduce the global detection efficiency threshold of the DI-QSS protocol employing the post-selection strategy by an average of 3.59$\%$, and reduce that of the basic DI-QSS protocol by an average of 6.79$\%$. The effect is relatively significant. When $F < 0.9$, the secret sharing rate is relatively small, rendering the reference value limited.

\subsection{Advantage Distillation and DI-QSS with the Post-Selection Strategy Based on Noise Pre-processing}
	
In noisy quantum channels, the global detection efficiency $\eta$ for photon signals is a secret parameter. Its theoretical model is jointly constructed from the efficiencies of three components: channel transmission, detector response, and optical path coupling. First, the transmission efficiency $\eta_{t}$ is primarily governed by channel attenuation, following the formula $\eta_t=10^{-\alpha d / 10}$, where $d$ is the fiber transmission distance and $\alpha$ is the loss coefficient. For standard optical fibers, $\alpha$ is typically taken as 0.2 dB/km. Second, the intrinsic detector efficiency $\eta_{d}$ is the quantum efficiency of the single-photon detector itself. Third, the photon-to-fiber coupling efficiency $\eta_{c}$ characterizes the efficiency with which photons couple from free space or a waveguide into the transmission fiber. The overall efficiency is determined by the product of these three components, i.e., $\eta = \eta_{t} \eta_{d} \eta_{c}$. Based on the current technological level of Superconducting Nanowire Single Photon Detectors (SNSPDs), detection efficiencies higher than 90$\%$, even reaching 98$\%$, have been achieved in the 1550 nm communication band\cite{zhang2017nbn,reddy2020superconducting,hao2024compact}. Therefore, in this simulation study, to evaluate the upper limit of system performance, we selected $\eta_{d}$=98$\%$ and $\eta_{c}$=99$\%$ as typical parameter values for setting and analysis.

\begin{figure}[htbp]
	\centering
	\includegraphics[width=0.9\columnwidth]{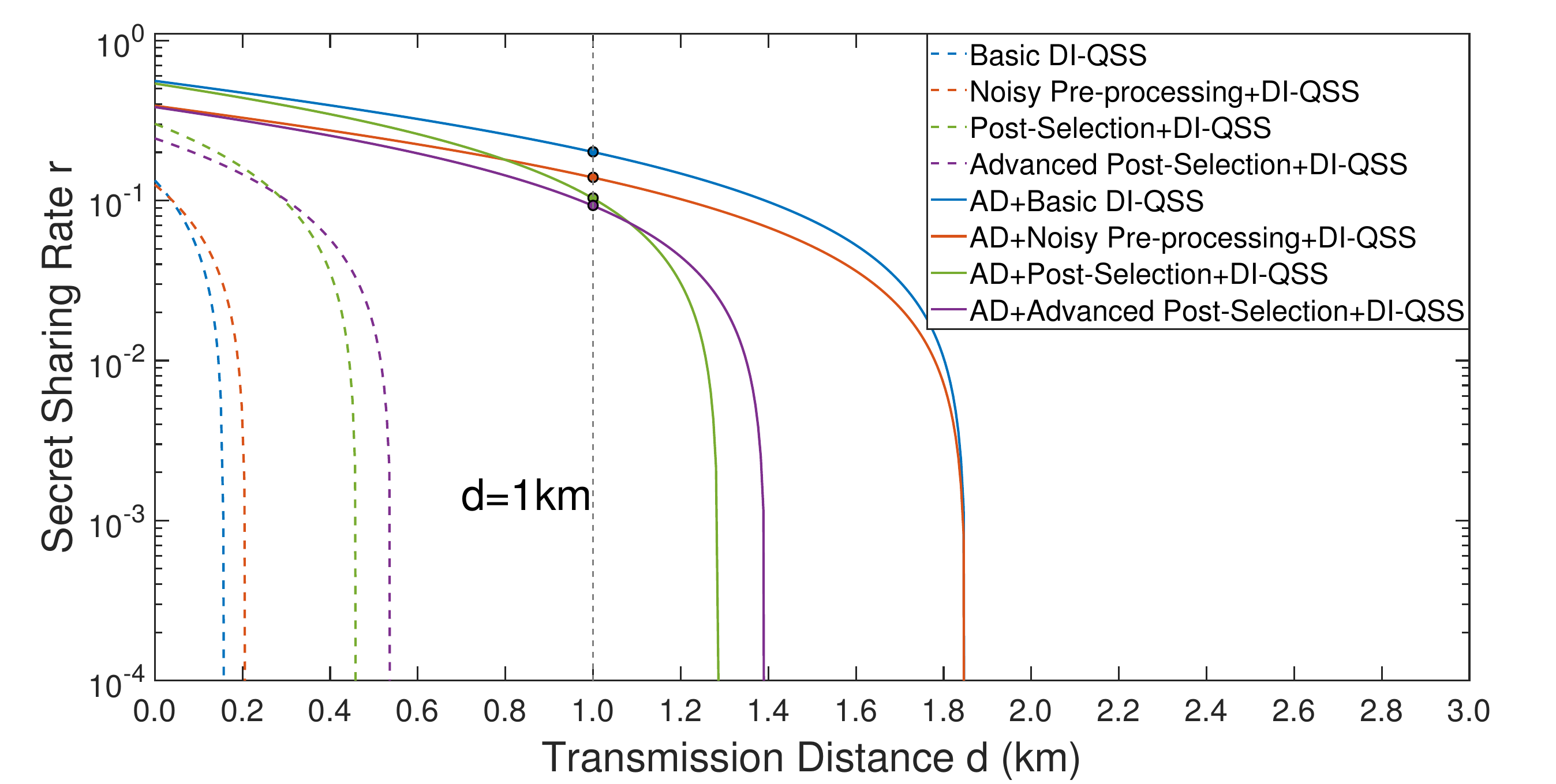}
	\caption{Plot of the secret sharing rate $r$ versus the transmission distance $d$ for the protocol under different strategies ($F=1,q=0.05$). This figure systematically compares the variation of the secret sharing rate $r$ with the transmission distance $d$ under four strategies, with (solid lines) and without (dashed lines) the advantage distillation technique.}
	\label{fig.13}
\end{figure}
\begin{figure}[htbp]
	\centering
	\includegraphics[width=0.9\columnwidth]{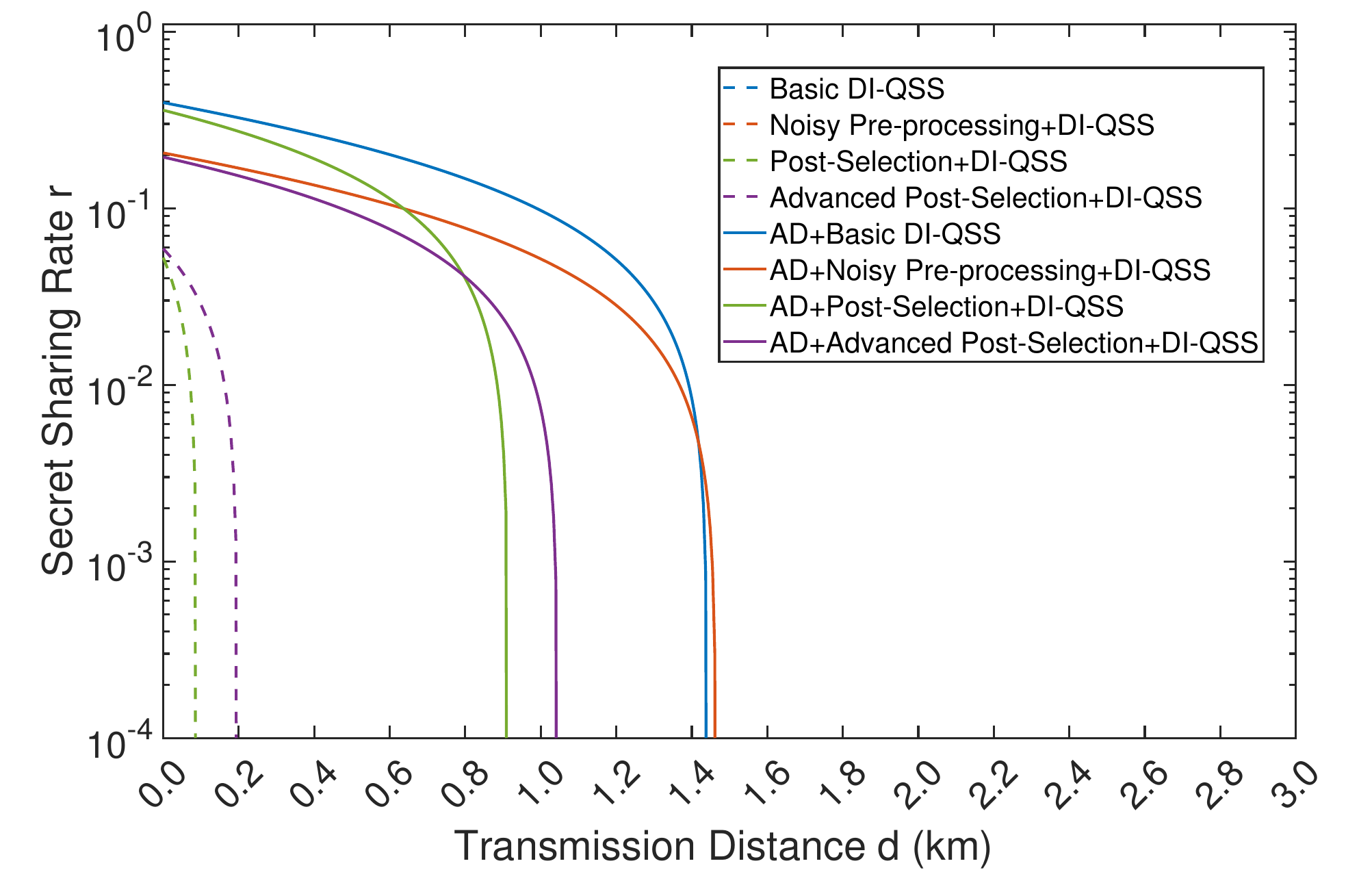}
	\caption{Plot of the secret sharing rate $r$ versus the transmission distance $d$ for the protocol under different strategies($F=0.95,q=0.05$).}
	\label{fig.14}
\end{figure}

The Fig.~\ref{fig.13} shows the variation of the secret sharing rate with the photon transmission distance for four scenarios before and after applying advantage distillation, when the fidelity $F=1$. The four scenarios are: the basic DI-QSS protocol, the noise pre-processing DI-QSS protocol ($q=0.05$), the post-selection DI-QSS protocol and the advanced post-selection DI-QSS protocol ($q=0.05$). It can be observed that under these four scenarios, the advantage distillation technique increases the maximum photon transmission distance of the protocols from 0.16 km to 1.85 km, from 0.20 km to 1.85 km, from 0.46 km to 1.28 km, and from 0.54 km to 1.39 km, respectively. In 2025, a DI-QSS protocol with advanced random key generation basis was proposed, which with equiprobable basis selection and a bit flip rate of $q=0.5$, increased the maximum photon transmission distance from 0.13 km to 0.71 km (source to user)\cite{zhang2025device}. Our protocol outperforms this one. Moreover, in the basic DI-QSS protocol and the DI-QSS protocol employing the noise pre-processing strategy, both combined with advantage distillation, the maximum secure communication distance between any two users is approximately 3.70 km, compared to 1.43 km for a protocol in 2025\cite{zhang2025device}. Furthermore, the basic protocol also exhibit the highest secret sharing rate. For example, when $d=1$ km, the secret sharing rate in this scenarios is 20.15$\%$, which is about twice that of the DI-QSS protocol based on the post-selection strategy with advantage distillation ($q=0.05$), where the secret sharing rate is $r$=10.37$\%$. 

In the practical scenario where the fidelity $F=0.95$, as shown in Fig.~\ref{fig.14}, the basic DI-QSS protocol and the DI-QSS protocol with noise pre-processing are already unable to achieve transmission communication. The optimization effect of advantage distillation on the noise pre-processing strategy is superior to that on the basic strategy. Both the basic strategy and the noise pre-processing strategy achieve simultaneously longer secure communication distances and higher secret sharing rates, making them promising choices for future applications.

\begin{table*}[tbp] % 注意这里的星号 *
	\setlength{\tabcolsep}{10pt}
	\centering
	\caption{Performance metrics comparison of different DI-QSS protocols}
	\label{tab:1} % 用于交叉引用
	\begin{tabular}{lcccc} % 定义你的列格式，这里示例为5列
		\toprule
		Protocol & $r$ & $\delta_{th}$ & $\eta_{th}$ & $d$(km) \\
		\midrule
		Basic DI-QSS & 0.234 & 10.17\% & 96.81\% & 0.16 \\
		Noise Pre-processing DI-QSS & 0.198 & 10.80\% & 96.59\% & 0.20 \\
		Post-selection DI-QSS & 0.357 & 7.15\% & 95.63\% & 0.46 \\
		Advanced Post-selection DI-QSS & 0.283 & 7.62\% & 95.28\% & 0.54 \\
		AD+Basic DI-QSS & 0.592 & 28.49\% & 89.72\% & 1.85 \\
		AD+Noise Pre-processing DI-QSS & 0.415 & 28.54\% & 89.70\% & 1.85 \\
		AD+Post-selection DI-QSS & 0.576 & 11.75\% & 92.06\% & 1.28 \\
		AD+Advanced Post-selection DI-QSS & 0.410 & 12.30\% & 91.61\% & 1.39 \\
		\bottomrule
	\end{tabular}
	\begin{tablenotes} % 可选，用于添加表格注释说明
		\small
		\item  \emph{Note.} The calculation premises for each performance indicator in the table are as follows. The secret sharing rate $r$ is calculated based on $F=0.98, q=0.05$ and $\eta$=0.98. The thresholds $\delta_{th}$ and $\eta_{th}$ are calculated based on $F=0.98$ and $q=0.05$. The maximum photon transmission distance $d$ is calculated based on $F=1, q=0.05$ and the condition that the detection efficiency $\eta$ satisfies the corresponding formula.
	\end{tablenotes}
\end{table*}

Under the conditions of fidelity $F=0.98$, qubit flip rate $q=0.05$, and global detection efficiency $\eta=0.98$, as is shown in TABLE~\ref{tab:1}. The secret sharing rate $r$ of the basic DI-QSS protocol is 0.234, which jumps to $r=0.592$ after applying advantage distillation, an improvement of over 2.5 times. For the noise pre-processing protocol, advantage distillation raises the secret sharing rate from 0.198 to 0.415. For the post-selection protocol, $r$ increases from 0.357 to 0.576. As for the advanced post-selection protocol, $r$ improves from 0.283 to 0.410. The data indicate that, regardless of whether the basic protocol already incorporates preliminary optimizations such as noise pre-processing or post-selection, advantage distillation can effectively extract reliable correlations for generating secret information, thereby markedly enhancing the final secret sharing rate.

From the perspective of protocol robustness, advantage distillation substantially relaxes the stringent requirements on practical system parameters. The four protocols originally require relatively high thresholds for global detection efficiency $\eta_{th}$ (0.95–0.97). After applying advantage distillation, the required $\eta_{th}$ decreases to approximately 0.89–0.92. More importantly, the noise tolerance $\delta_{th}$ of the protocols is greatly improved with advantage distillation. Specifically, the $\delta_{th}$ of the basic protocol improves from 10.17$\%$ to 28.49$\%$, the noise pre-processing protocol from 10.80$\%$ to 28.54$\%$, the post-selection protocol from 7.15$\%$ to 11.75$\%$, and the advanced post-selection protocol from 7.62$\%$ to 12.30$\%$. This further demonstrates that advantage distillation provides greater optimization for protocols that do not employ post-selection strategy. For the DI-QSS protocol with advanced random key generation basis, when the basis is chosen with equal probability and the qubit flip rate $q$ is 0.5, it improves the $\delta_{th}$ from 7.15$\%$ to 9.23$\%$ and reduces the $\eta_{th}$ from 96.32$\%$ to 93.41$\%$\cite{zhang2025device}. The optimization achieved by our protocol still outperforms that of the aforementioned protocol. Moreover, advantage distillation significantly reduces the actual QBER of the protocols. This means advantage distillation effectively filters out errors introduced by channel noise and attacks, enabling the protocols to withstand higher levels of noise interference. Consequently, the protocols can remain operational even in practical systems with lower photon collection efficiency and higher transmission loss, thereby greatly enhancing their feasibility for real-world deployment.

In summary, the numerical analysis in this section clearly demonstrates that advantage distillation is a key and highly effective technique for improving the practical performance of DI-QSS protocols. It not only directly enhances the secret sharing rate but also fundamentally strengthens the robustness of the protocols when facing real-world challenges such as high loss and high noise. This dual-optimization effect implies that DI-QSS protocols equipped with an advantage distillation module can achieve stable and secure secret sharing even over longer-distance, higher-loss quantum links, or on experimental platforms with limited detector efficiency and strong noise backgrounds. This provides a more feasible protocol design for building large-scale, high-fault-tolerant quantum-secure networks, representing a crucial step toward advancing DI-QSS from theoretical verification to practical application. 

\section{Conclusion}\label{five}
This paper addresses the practical bottlenecks of device-independent quantum secret sharing (DI-QSS) protocols\cite{zhang2024device}, such as low noise tolerance, stringent requirements for global detection efficiency, and limited secure distance caused by channel loss and noise, by proposing a systematic optimization scheme based on the advantage distillation technique. We have successfully extended the advantage distillation framework to the three-party DI-QSS scenario. By designing a data filtering and verification process adapted to three-party interaction, we have developed a post-processing method capable of significantly enhancing the robustness of the protocol. This method achieves a substantial improvement in noise tolerance at the cost of a slight reduction in the data rate, effectively alleviating the overly stringent experimental requirements of the original protocol. We systematically applied this advantage distillation technique to the basic DI-QSS protocol and its three active improvement strategies, deriving analytical expressions for the effective QBER and the lower bound of the secret sharing rate for each after incorporating advantage distillation, thereby establishing a complete theoretical model for performance optimization. Comprehensive numerical simulations have verified the operational mechanism and optimization effects of the scheme. The results demonstrate that the advantage distillation technique can generally enhance the protocol's noise tolerance, lower the global detection efficiency threshold, and significantly extend the maximum secure communication distance.

Further analysis reveals that the difference in optimization effectiveness depends on the ``source of advantage". Performance improvement is most pronounced in the basic protocol and the noise preprocessing strategy, where initial quantum correlations are high. In contrast, the optimization effect is relatively weaker in the post-selection strategy, which introduces substantial classical random noise. In the future, we can consider more diversified optimization strategies to compensate for the limitations of a single approach. For instance, the combination of the advantage distillation technique\cite{tan2020advantage,liu2023mode,zhu2023reference,du2025advantage,stasiuk2025quantum} with a random key generation basis strategy\cite{schwonnek2021device,masini2022simple,zhang2025device} may yield promising results.

\vspace{1em} % 可根据需要调整数值
\section*{Data Availability}
The data that support the findings of this article are not publicly available. The data are available from the authors upon reasonable request.

\appendix

\section{}\label{A}

The proof of Eq.~\eqref{eq.5} is given below. The density matrix for the tripartite qubit system of Alice, Bob and Charlie is given as follows:
\begin{equation}
	\rho_{ABC} = \eta^{3} \left[ F |\phi_{1}\rangle \langle\phi_{1}| + \frac{1-F}{8} I \right],\ \ 
	\sum_{i=1}^{8} |\phi_{i}\rangle \langle\phi_{i}| = I
	\tag{A1} %将公式编号覆盖为自定义的"A1"
	\label{eq:A1} %允许通过 \ref{eq:A1}引用此公式
\end{equation}
Among them, the eight GHZ basis states are given as follows:
\begin{align*}
	\left|\phi_{1}\right\rangle &= \frac{1}{\sqrt{2}}(|H\rangle|H\rangle|H\rangle+|V\rangle|V\rangle|V\rangle)\\ &= \frac{1}{2}(|+++\rangle+|+--\rangle+|-+-\rangle+|--+\rangle) \\
	\left|\phi_{2}\right\rangle &= \frac{1}{\sqrt{2}}(|H\rangle|H\rangle|H\rangle-|V\rangle|V\rangle|V\rangle)\\ &= \frac{1}{2}(|++-\rangle+|+-+\rangle+|-++\rangle+|---\rangle) \\
	\left|\phi_{3}\right\rangle &= \frac{1}{\sqrt{2}}(|V\rangle|H\rangle|H\rangle+|H\rangle|V\rangle|V\rangle)\\ &= \frac{1}{2}(|+++\rangle+|+--\rangle-|-+-\rangle-|--+\rangle) \\
	\left|\phi_{4}\right\rangle &= \frac{1}{\sqrt{2}}(|V\rangle|H\rangle|H\rangle-|H\rangle|V\rangle|V\rangle)\\ &= \frac{1}{2}(|++-\rangle+|+-+\rangle-|-++\rangle-|---\rangle) \\
	\left|\phi_{5}\right\rangle &= \frac{1}{\sqrt{2}}(|H\rangle|V\rangle|H\rangle+|V\rangle|H\rangle|V\rangle)\\ &= \frac{1}{2}(|+++\rangle-|+--\rangle+|-+-\rangle-|--+\rangle) \\
	\left|\phi_{6}\right\rangle &= \frac{1}{\sqrt{2}}(|H\rangle|V\rangle|H\rangle-|V\rangle|H\rangle|V\rangle)\\ &= \frac{1}{2}(|++-\rangle-|+-+\rangle+|-++\rangle-|---\rangle) \\
	\left|\phi_{7}\right\rangle &= \frac{1}{\sqrt{2}}(|H\rangle|H\rangle|V\rangle+|V\rangle|V\rangle|H\rangle)\\ &= \frac{1}{2}(|+++\rangle-|+--\rangle-|-+-\rangle+|--+\rangle) \\
	\left|\phi_{8}\right\rangle &= \frac{1}{\sqrt{2}}(|H\rangle|H\rangle|V\rangle-|V\rangle|V\rangle|H\rangle)\\ &= \frac{1}{2}(-|++-\rangle+|+-+\rangle+|-++\rangle-|---\rangle)
\end{align*}
Let the measurement basis $\{ |{+++}\rangle,\ |{+--}\rangle,\ |{-+-}\rangle,$\\
$|{--+}\rangle,|{++-}\rangle,\ |{+-+}\rangle,\ |{-++}\rangle,\ |{---}\rangle \}$ be used to measure the quantum state $\rho_{ABC}$. The probability of obtaining the measurement outcome $|ijk\rangle$ is given by $P_{ijk} = \langle ijk | \rho_{ABC} | ijk \rangle,\ i,j,k \in \{+,-\}$, which $``+"\rightarrow``0", ``-"\rightarrow``1"$. Therefore,
\begin{align*}
		P_{000} &= \bigl\langle +++ \bigr| \eta^{3} \bigl[ F|\phi_{1}\rangle\langle\phi_{1}| 
		+ \tfrac{1-F}{8} I \bigr] \bigl| +++ \bigr\rangle \\
		&= \eta^{3}F \langle +++   |\phi_{1}\rangle\langle\phi_{1}| +++ \rangle + \tfrac{1}{8} \eta^{3} (1-F) \\
		&= \frac{1}{8}(1+F)\eta^3
		\label{A2}
		\tag{A2}
\end{align*}

Similarly, the probabilities for the other measurement outcomes are:
\begin{align*}
	p_{011} &= \langle +-- | \rho_{ABC} | +-- \rangle = \frac{1}{8}(1+F)\eta^{3} && \\
	p_{101} &= \langle -+- | \rho_{ABC} | -+- \rangle = \frac{1}{8}(1+F)\eta^{3} && \\
	p_{110} &= \langle --+ | \rho_{ABC} | --+ \rangle = \frac{1}{8}(1+F)\eta^{3} && \\
	p_{001} &= \langle ++- | \rho_{ABC} | ++- \rangle = \frac{1}{8}(1-F)\eta^{3} && \\
	p_{010} &= \langle +-+ | \rho_{ABC} | +-+ \rangle = \frac{1}{8}(1-F)\eta^{3} && \\
	\\
	p_{100} &= \langle -++ | \rho_{ABC} | -++ \rangle = \frac{1}{8}(1-F)\eta^{3} && \\	
	p_{111} &= \langle --- | \rho_{ABC} | --- \rangle = \frac{1}{8}(1-F)\eta^{3} &&
\end{align*}

When the measurement outcomes $a, b, c$ are 000, 011, 101, 110, the tripartite result satisfies $a \oplus b \oplus c=0$; when the measurement outcomes $a, b, c$ are 001, 010, 100, 111, the tripartite result does not satisfy $a \oplus b \oplus c=0$, which corresponds to erroneous cases. The bit error rate before advantage distillation is given by
\begin{align*}
	\delta = p_{010} + p_{100} + p_{001} + p_{111} =1- \frac{1}{2}(1+F)\eta^{3}
	\tag{A3}
\end{align*}
Thus, the proof of Eq.~\eqref{eq.6} is completed.

Now we calculate the bit error rate after the three parties perform advantage distillation with a block length of 2 bits. Suppose Alice, Bob and Charlie share two triples of bits: $\left(a_{1}, b_{1},c_{1}\right),\left(a_{2}, b_{2},c_{2}\right)\in\{0,1\}^{3}$. According to the rule of tripartite advantage distillation, the condition is satisfied if and only if $ \left(a_{1} \oplus b_{1} \oplus c_{1}=0, a_{2} \oplus b_{2} \oplus c_{2}=0\right) $
\text{or} $ \left(a_{1} \oplus b_{1} \oplus c_{1}=1, a_{2} \oplus b_{2} \oplus c_{2}=1\right) $ holds, the triple $\left(a_{1}, b_{1},c_{1}\right)$ is retained; otherwise, both blocks are discarded including those with photon loss. The total probability of retaining the data under all possible scenarios after advantage distillation is therefore:
\begin{align*}
	p_{\text{retained}} &= \left(p_{000} + p_{011} + p_{101} + p_{110}\right)^{2}\\
	&+ \left(p_{010} + p_{100} + p_{001} + p_{111}\right)^{2} \\
	&= \frac{1}{4}\eta^{6}\left[(1+F)^{2} + (1-F)^{2}\right]
	\tag{A4}
\end{align*}
The probability of bit errors in the surviving blocks after advantage distillation is:
\begin{align*}
	\delta_{ad} &=\frac{\left(p_{010}+p_{100}+p_{001}+p_{111}\right)^{2}}{p_{\text{retained}}} \\
	&=\frac{(1-F)^{2}}{(1+F)^{2}+(1-F)^{2}}
	\tag{A5}
\end{align*}
Thus, the proof of Eq.~\eqref{eq.5} is completed.

\bibliographystyle{apsrev4-2} 
\bibliography{reference}% Produces the bibliography via BibTeX.

\end{document}